\documentstyle[sprocl]{article}
\input{psfig}
\font\FermiPPTfont=cmssbx10 scaled 1440
\font\FermiSmallfont=cmssq8 scaled 1200

\def\FNALpptheadhigh#1#2{
\null \vskip -1.5truein
\centerline{\hbox to 7.5truein {
\vbox to 1in{\vfill 
             \hbox{\psfig{figure=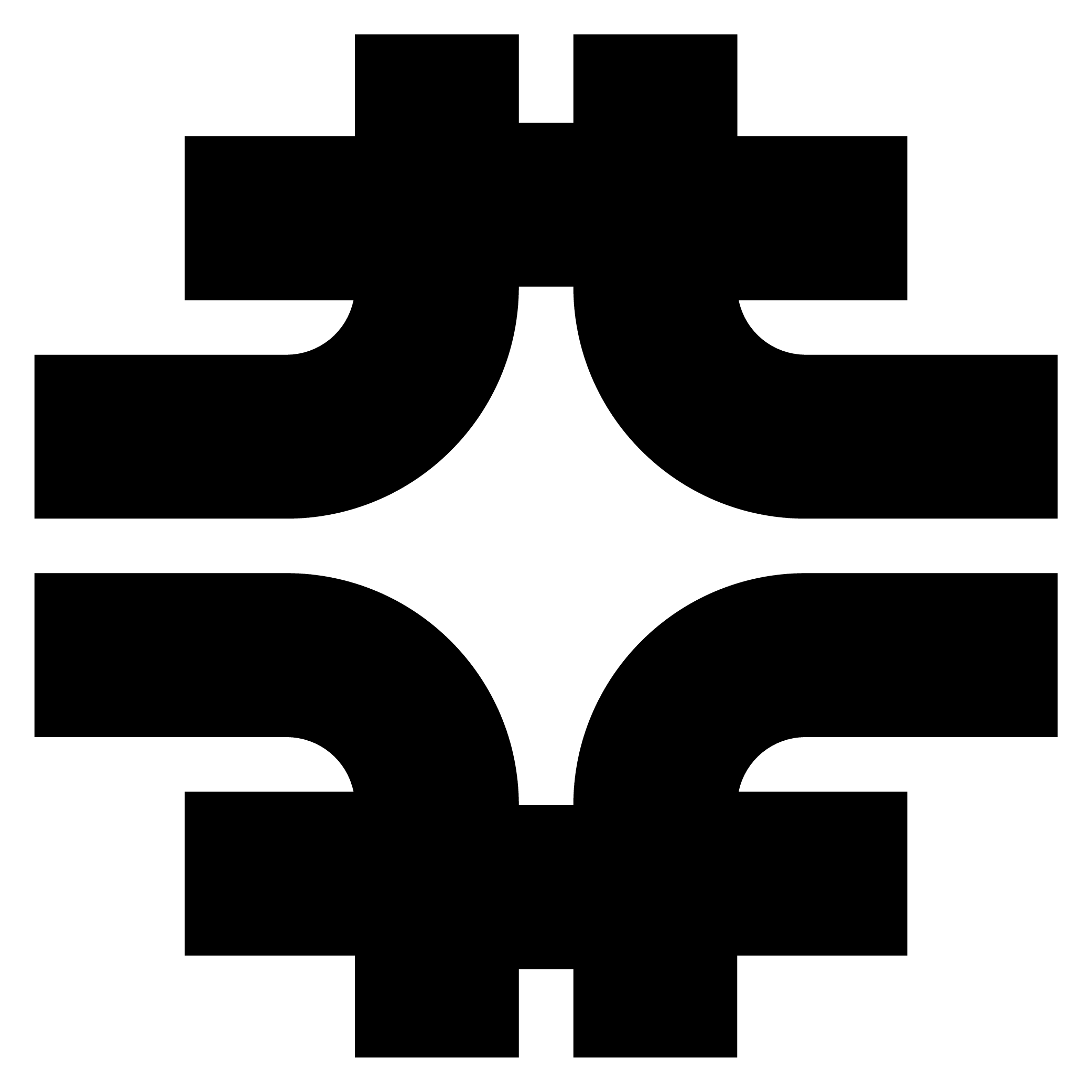,height=1.5cm,clip=}} 
             \vfill }
\hskip 1em
\vbox to 1in{\vfill
             \hbox{{\FermiPPTfont Fermi National Accelerator Laboratory}}
             \vfill}
\hfill
\vbox to 1in {\vfill \FermiSmallfont
              \hbox{#1}
              \hbox{#2}
              \vfill}
}}}
\def\Journal#1#2#3#4{{#1} {\bf #2}, #3 (#4)}

\def\etai{{\eta_\rmi}}
\def\etaf{{\eta_\rmf}}

\def\spar{{\s\parallel}}
\def\lmin{{l_{\rm min}}}
\def\lmax{{l_{\rm max}}}

\def\Ylm{{Y_{\s(l,m)}}}
\def\alm{{a_{\s(l,m)}}}
\def\tildealm{{\widetilde{a}_{\s(l,m)}}}
\def\Deltalm{{\Delta_{\s(l,m)}}}
\def\pmb#1{\setbox0=\hbox{#1}%
  \kern-.025em\copy0\kern-\wd0
  \kern.05em\copy0\kern-\wd0
  \kern-.025em\raise.0433em\box0 }
\def\timedate{ {\tt
\count215=\time \divide\count215 by60  \number\count215
\multiply\count215 by-60 \advance \count215 by\time :\number\count215 \space
\number\day\space
\ifcase\month\or January\or February\or March\or April\or May\or June\or July
\or August\or September\or October\or November\or December\fi\space\number\year
}}
\def \etal {{\it et al.} }

\def\s {\scriptscriptstyle}
\def\ccdot{{\hskip-0.7pt\cdot\hskip-0.7pt}}

\def\sqr#1#2{{\vcenter{\hrule height.#2pt
              \hbox{\vrule width.#2pt height#1pt \kern#1pt \vrule width.#2pt}
              \hrule height.#2pt}}}

\def\mathrelfun#1#2{\lower3.6pt\vbox{\baselineskip0pt\lineskip.9pt
  \ialign{$\mathsurround=0pt#1\hfil##\hfil$\crcr#2\crcr\sim\crcr}}}
\def\simlt{\mathrel{\mathpalette\mathrelfun <}}
\def\simgt{\mathrel{\mathpalette\mathrelfun >}}

\def\rmf {{\rm f}}

\def\rmi {{\rm i}}

\def\rmk {{\rm k}}

\def\rmI {{\rm I}}

\def\rmS {{\rm S}}
\def\rmT {{\rm T}}

\def\rmV {{\rm V}}

\def\bfk {{\bf k}}

\def\bfm {{\bf m}}
\def\bfn {{\bf n}}

\def\bfx {{\bf x}}
\def\bfy {{\bf y}}
\def\bfz {{\bf z}}

\def\bfI {{\bf I}}

\def\bfN {{\bf N}}

\def\hatbfk  {{\hat\bfk}}

\def\hatbfm  {{\hat\bfm}}
\def\hatbfn  {{\hat\bfn}}

\def\hatbfx  {{\hat\bfx}}
\def\hatbfy  {{\hat\bfy}}
\def\hatbfz  {{\hat\bfz}}

\def\eV  {{\rm \hbox{e\kern-0.14em V}}}
\def\keV {{\rm \hbox{ke\kern-0.14em V}}}
\def\MeV {{\rm \hbox{Me\kern-0.14em V}}}
\def\GeV {{\rm \hbox{Ge\kern-0.14em V}}}

\begin{document}
\FNALpptheadhigh{NASA/Fermilab Astrophysics Center}{Fermilab-Conf-97/133-A}
\title{ON THE COMPUTATION OF CMBR ANISOTROPIES FROM SIMULATIONS OF TOPOLOGICAL
DEFECTS}
\author{A. Stebbins and S. Dodelson}
\address{NASA/Fermilab Astrophysics Center, \\
Fermilab, Box 500, Batavia, IL 60510, USA}
\maketitle\abstracts{
Techniques for computing the CMBR anisotropy from simulations of topological
defects are discussed with an eye to getting as much information from a
simulation as possible. Here we consider the practical details of which sums
and multiplications to do and how many terms there are.}

	Probably the most fruitful field of cosmological observation in the
past few years comes from observations of the anisotropies of the Cosmic
Microwave Background Radiation (CMBR).  One of the first and the most dramatic
detections was made by the DMR on the COBE satellite over 5 years ago.  New
results continue to pour in and we can expect this continue to at least until
the Planck experiment provides us with an exquisitely detailed all-sky map in
about a decade.  Theoretical modeling of the data and their interpretation in
terms of cosmogenic theories has generally been able to keep pace with these
observations but we can expect the observational data to become increasingly
more precise.  There are models such as inflation where the statistical
properties of the anisotropy pattern is well described as being statistically
isotropic Gaussian random noise where the parameters describing the Gaussian
distribution can be computed by solving linear sets of ordinary differential
equations.  For such models a complete statistical description is fairly easy
to obtain and the predictions are routinely made made with better than 1\%
accuracy.  In another class of theories, where the inhomogeneities are seeded
by topological defects, it is not clear whether we will ever obtain this kind
of accuracy.  Of course this may be a moot point if the predictions for these
models, such as they are, do not fit observations.  In the paper we will
discuss some of the issues involved with computing CMBR anisotropies for defect
models.

	One of the fundamental differences between Gaussian models and defect
models is that the dynamics of the former are described by linear equations
while that the latter by nonlinear equations.  Linearity combined with the
assumed statistical homogeneity of the universe guarantee that one may
decompose the cosmological perturbations into eigenmodes, which are
eigenfunctions of the generators of the isometries of the space-time, and that
each eigenfunction will evolve independently.  So, for example in a flat
Friedmann-Robertson-Walker (FRW) spacetime, the eigenmodes could be Fourier
modes which are eigenfunctions of the generator of translation. While one might
have to solve the linear equations for a Fourier mode numerically, one need
only solve the equation once for each value of $|\bfk|$. For a given type of
inhomogeneity, e.g. adiabatic or isocurvature, the only freedom in the initial
condition of a given mode is the amplitude, and, since the equations are
linear, solutions scale linearly with the initial amplitude.

	For topological defects the equations are non-linear so different
eigenmodes are coupled and one does not know the evolution of one mode without
taking into account the evolution of {\it all} the other.  It is usually
simpler to think about topological defects in real space rather than the space
of eigenmodes ($k$-space).  One reason is that equations describing the
evolution of the defects are spatially local while they are highly non-local in
$k$-space.  Another reason is that causality guarantees that the range
interaction does not extend beyond a certain distance, the causal horizon,
while the range of interaction in $k$-space is unbounded. Causality allows one
to consider a finite patch of the universe and ignore what goes on outside of
that patch, at least for points more than a horizon distance away from the
boundary of the patch.

	In order to determine the likelihood that the observed inhomogeneities
in the universe could be explained by topological defects one would need a
description of the statistical ensemble of possible topological defects
configurations and their evolution.  In general we cannot even find analytic
solution to describe the evolution of specific configurations so the way one
tries to determine the statistical properties is through numerical experiment.
Namely one generates a realization of the defect configuration at some early
time and evolves it and the matter surrounding it according to their equations
of motion.  One might then try another configuration, and so on, until one has
a fair statistical sample.  It is generally believed that the statistical
properties of the defects are {\it ergodic} so that one may also obtain a fair
sample by considering a large enough volume of a single realization.  One need
only consider the observations predicted for observers situated in different
locations in the simulation volume.  Since one needs to build up a fair sample
it is important to obtain as many realizations of the anisotropy as possible.

\section{Geometry of Computation}

\begin{figure}
\psfig{figure=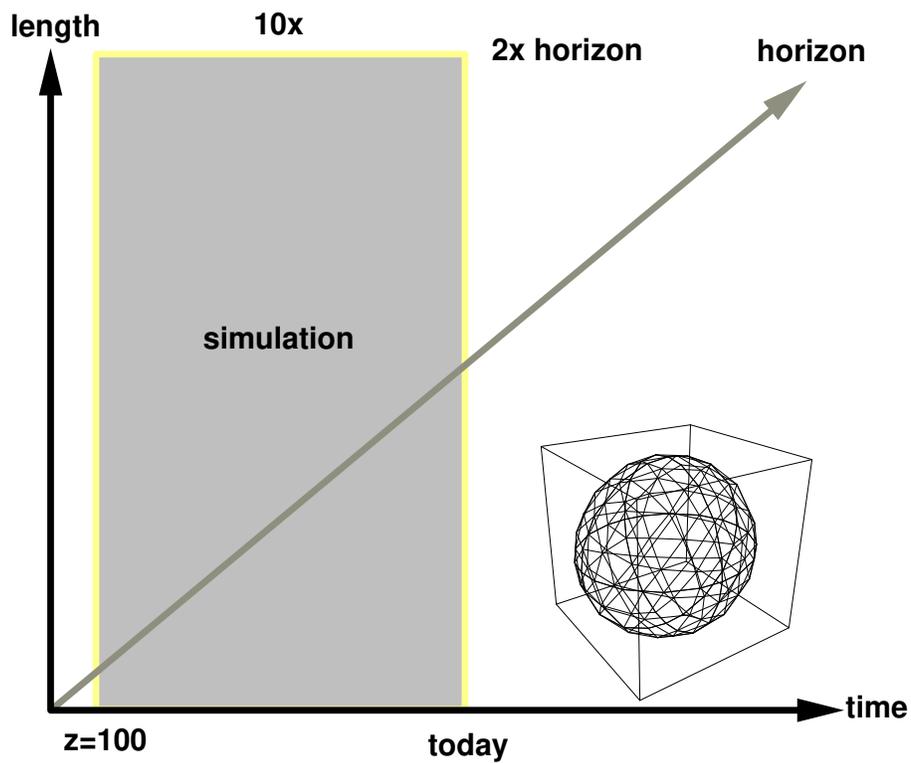,height=10cm}
\caption{Here is shown the geometry for large angle anisotropy computations.
In the inset the cube represents the simulation volume while the sphere gives
the observer's horizon, the volume which contributes to the anisotropy pattern
of a given observer. The shaded on the graph gives the range of temporal and
length scales resolved by the simulation.  Generally the most important missing
scales are on the left, i.e. from early times. Some numbers appropriate for
studying COBE scale anisotropies are given.}
\label{fig:LargeAngle}
\end{figure}

	Numerous authors have used simulations of defects to produce
realizations of anisotropy patterns from defects
\cite{BBS88,BennettRhie93,DHZ93,PST94,PiInSky94,PS95,DZ96,Turok96,ACSSV96}.
Some of these have been used to examine large angle anisotropies
\cite{BennettRhie93,DHZ93,PST94,PS95,DZ96,ACSSV96} while others have been used
to examine anisotropies on smaller angular scales
\cite{BBS88,PiInSky94,Turok96}.  The geometry used for these large angle {\it
numerical experiments} is shown in fig~\ref{fig:LargeAngle}.  One must evolve
the defects in a very large simulation volume, as big as the present horizon
and ideally one would like to evolve from time of recombination.  Usually one
takes periodic boundary conditions for the defect configuration.  However this
periodicity does not really matter - since for most methods of laying down the
defects one can show that the defect configuration in any given horizon volume
is drawn from the same distribution as if one were to have simulated and
infinite volume.  The cubical box with periodic boundary condition are only
easily implemented in the case of spatially flat FRW cosmology.  Doing this
computation for an open universe, as was done in by Pen and Spergel\cite{PS95},
is actually a more difficult problem in many respects.  For simplicity we will
assume a flat FRW cosmology, which allows us to make use of the Fourier series
representations of distributions in the simulation box, i.e.
\begin{equation}
F(\bfx)=\sum_\bfk \widetilde{F}(\bfk)\,e^{i\bfk\ccdot\bfx}\ .
\end{equation}
We will specifically be considering a a cubical simulation volume with 
$N\times N\times N$ spatial resolution elements.

\begin{figure}
\psfig{figure=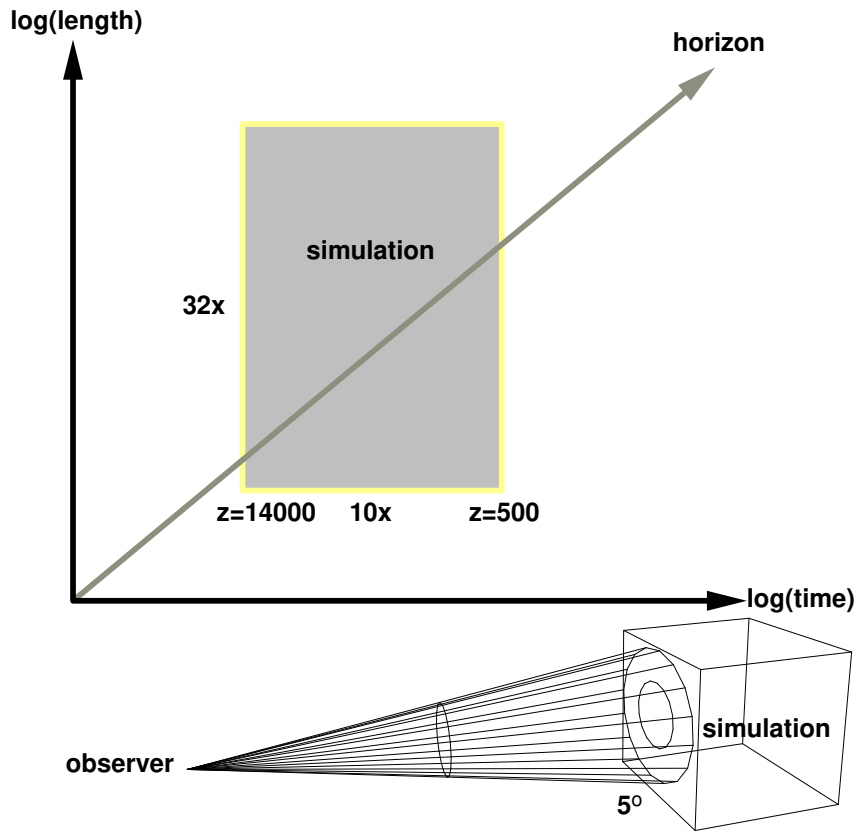,height=12cm}
\caption{Here is shown the geometry for a small-angle anisotropy computations.
The cube represents the simulation volume and the cone the photons from the
simulation converging toward the observer. The shaded region on the graph gives
the scales resolved by the simulation. The most important missing scales are on
the right and left, i.e. early times and late times. Some numbers appropriate
for studying degree scale anisotropy are given.}
\label{fig:SmallAngle}
\end{figure}

\section{Dynamic Range}

	Given these general considerations let us proceed to consider specific
issues associated with the computation of MBR anisotropies from seeds in a box
of a fixed comoving size, $L$.  All numerical simulations have a finite
dynamical range in both spatial and temporal scales.  There will be a finite
spatial resolution on small scales and the size of the box itself will limit
the largest accessible scale.  Similarly there will be finite temporal
resolution, usually related to spatial resolution by the Courant condition
$\Delta t\simlt\Delta x/c$.  Here $c$ is the speed-of-light which characterizes
the dynamics of most defects.  For this reason the coherence length of the
defects will scale with the causal horizon, so that in order to properly
represent the defect configuration the finite spatial resolution leads to a
minimum starting time: $t_\rmi\gg\Delta x$.  This large characteristic velocity
of the defects also means that the defect configuration will start to ``see''
the finite size of the box when the causal horizon reaches the box size.  It is
desirable to end the simulations before this happens, i.e. $t_\rmf\simlt L/c$.
These considerations indicate that the number of temporal resolution elements
is $\sim N$.

From a box at least as large as twice the horizon one can study the very
largest angle anisotropies since, as illustrated in fig~\ref{fig:LargeAngle},
one does not see the periodicity of the universe. The edge of ones horizon in
one direction nearly touches the edge in the opposite direction, but one is
seeing this far only at the earliest times when the (causal) correlation length
is arbitrarily small, so there need be no effect of periodicity at all.  The
finite spatial resolution in such a simulation will limit the angular
resolution.  If most of the anisotropy is generated at large redshifts then the
angular resolution is $\Delta\theta\sim\Delta x/(L/2)\sim2/N$, which
corresponds to a largest available angular wavenumber $\lmax\sim N/2$.

	For simulations used to compute smaller angle anisotropy, as
illustrated in fig~\ref{fig:SmallAngle}, the effects of periodicity are much
more severe as one can ``see'' beyond the simulation volume.  One is not really
simulating a large universe as we live in, but rather a truly small and
periodic one.  In such a universe one would see the pattern of anisotropy to be
roughly periodic on the sky.  What one is hoping is that the statistical
properties of each of these patches are similar to those of a similar sized
patch in a non-periodic universe.  The angle subtended by the simulation box
gives you an upper limit on the angular scale you are probing, while the
spatial resolution gives a lower limit.  Clearly the range of scales probed is
roughly the same as for the large-angle case: $\lmax/\lmin\sim N$.  Note
however that one is not assured that one has the complete picture for these
scales since one also is limited in temporal scales as well (see
fig~\ref{fig:SmallAngle}).  One could, in principle, piece together all scales
by combining simulations with different box sizes and different temporal
ranges.  Being able to do so presupposes that the different angular and
temporal scales are essentially uncorrelated. This is liable to be less true in
non-Gaussian models.  For example in string models, even long strings at late
times which subtend a very large angle, do also exhibit temperature
discontinuities which will have an effect on all angular scales.  However the
discontinuities from these longer strings are rare and do not contribute
greatly to the smaller scale anisotropies when compared to the much more
numerous smaller angle strings.  The assumption of independence of scales is a
good approximation but not exactly correct.  Note also that the finite box size
not only limits the access to large spatial scales but also to large temporal
scales.  Once the horizon sufficiently exceeds the box size, the seeds will
behave differently than they would in an infinite universe.  Thus we must also
presuppose that there are no long term temporal correlations which will be
different in periodized box than in an infinite box.  Generally we find that
seeds only produce significant perturbations on scales comparable to the
horizon size and thus any long term temporal correlations which might exist
cannot not have much effect on the anisotropies.

\section{The Brightness Distribution}

	Here we would first like to point out just how much information is
required to describe the photon distribution present in a simulation.  Recall
that in a Gaussian models the statistical distribution of the anisotropies
observed by any given observer is determined completely by the
ensemble-averaged angular power spectrum, i.e. the $C_l$'s.  For angular
resolution characterized by $\lmax$ one only needs $\lmax$ real numbers.  Note
that the $C_l$'s do not give the spatial correlations in the anisotropy pattern
but, as this is not observable in practice, these spatial correlations are not
of interest.  For a non-Gaussian model where one relies on numerical experiment
one really wants to use the distribution function of anisotropy patterns for
all observers in the simulation.  If one has $N^3$ spatial resolution elements
and $(\lmax+1)^2$ angular resolution elements then one needs naively
$N^3(\lmax+1)^2$ real numbers to describe the brightness
pattern.\footnote{We use $(\lmax+1)^2$ for the number of of angular
resolution elements which is really only appropriate for a large-angle
simulation, although the number of independent lines-of-sight is $\sim N^2$ for
both small and large-angle simulations - and the size of the computation is the
same in both cases.}
So, for example, we could describe the brightness pattern as
\begin{equation}
{\Delta T\over T}(\hatbfn,\bfx)=
\sum_{l=0}^\lmax \sum_{m=-l}^l\alm(\bfx)\,\Ylm(\hatbfn;\hatbfz)\ .
\label{YlmDecomposition}
\end{equation}
where the $\alm$'s are the $N^3(\lmax+1)^2$ real numbers (recall that 
$a_{\s(l,-m)}=(-1)^m\alm^*$ and there are thus $2l+1$ real numbers for each
$l$).  Since typically $\lmax\sim N$ this is $N^5$ numbers - which is a rather
formidable amount of data.  Note that there is some information about the
uninteresting spatial correlations in these numbers.  In fact one can use the
spatial correlations to significantly reduce the amount of information needed
to describe the brightness distribution.

	The Fourier transform of eq~(\ref{YlmDecomposition}) is 
\begin{equation}
{\Delta T\over T}(\hatbfn,\bfx)=\sum_\bfk \sum_{l=0}^\lmax\sum_{m=-l}^l
                 \tildealm(\bfk)\,\Ylm(\hatbfn;\hatbfz)\,e^{i\bfk\ccdot\bfx}\ .
\label{FourierYlm}
\end{equation}
which still contains $N^3(\lmax+1)^2$ real numbers.  Here $\hatbfn$ gives the
direction in which the brightness is measured.  The 2nd argument of $\Ylm$,
$\hatbfz$, indicates the North pole of the spherical polar coordinates used in
the spherical harmonic decomposition.  In this case the $m$ decomposition has
no particular physical significance.  If however we rotate the spherical polar
coordinates, differently for each $\bfk$, such that the North pole is the
direction of $\bfk$, we obtain a more interesting decomposition:
\begin{equation}
{\Delta T\over T}(\hatbfn,\bfx)=\sum_\bfk \sum_{l=0}^\lmax\sum_{m=-l}^l
                 \Deltalm(\bfk)\,\Ylm(\hatbfn;\hatbfk)\,e^{i\bfk\ccdot\bfx}
.\label{FourierHelicity}
\end{equation}
Mathematically this rotation is given by
\begin{equation}
\tildealm(\bfk)=\sum_{m'=-l}^l D^l_{m,m'}(\hatbfz,\hatbfk)\,\Deltalm(\bfk)\ ,
\label{WignerRotation}
\end{equation}
where the $D^l_{m,m'}$ are Wigner rotation matrices defined by
\begin{equation}
D^l_{m,m'}(\hatbfz,\hatbfk)=
\int d^2\hatbfn\,Y^*_{\s(l,m)}(\hatbfn;\hatbfz)\,Y_{\s(l,m')}(\hatbfn;\hatbfk)
\label{WignerMatrices}
\end{equation}
which obey the sum rule
\begin{equation}
\sum_{m=-l}^l D^l_{   mm_1}(\hatbfz,\hatbfk)\,
                D^{l*}_{mm_2}(\hatbfz,\hatbfk)=\delta_{m_1m_2}
.\label{WignerSumRule}
\end{equation}
The $m$ decomposition of eq~(\ref{FourierHelicity}) is useful is because the
different $m$'s correspond to the different {\it helicities} of the brightness
pattern.  In linear theory (i.e. small perturbations) the different helicities
of the brightness pattern only couple to different components of the
gravitational field. In particular $m=0$ couples to {\it scalar} perturbations,
$m=\pm1$ to {\it vector} perturbations, and $m=\pm2$ to {\it tensor}
perturbations.  Helicities with $|m|>2$ are not coupled to the gravitational
field at all.   Any brightness anisotropy with $|m|>2$ helicities, with no
gravitational ``forcing terms'', would be damped to zero by scattering of the
photons.  Thus we expect that the brightness pattern in our universe to only
have helicities with $|m|\le2$, i.e. scalar, vector, and tensor modes.  It
would indeed be interesting, as a check of this notion, to decompose the
brightness pattern in our patch of the universe into helicities.  However this
requires a knowledge of the spatial pattern of anisotropies, and we cannot
expect to measure this anytime soon.  Nevertheless, in a simulation we do know
the spatial pattern, and we may use the helicity decomposition, as a form of
data compression by ignoring the $\Deltalm$'s with $|m|>2$ which we know {\it a
priori} to be zero.  By doing so we have reduced the amount of data required to
describe the brightness pattern to $5\lmax N^3\sim N^4$, i.e. reducing the
dimensionality by one.  The special ``helicity state'' of the CMBR brightness
pattern has long been understood\cite{AbbottSchaefer86} and is implicit in most
anisotropy computations, but the transformation between helicities and the
spatial brightness distribution is rarely used as it is not needed in Gaussian
theories.

	In Gaussian theories the brightness pattern is described by the $C_l$'s
and in non-Gaussian theories the $C_l$'s still exist but do not give the entire
picture.  Since the $C_l$'s give the expectation value of the $|\alm|^2$
averaged over all realizations one cannot determine their values from a single
simulation.  However, just as we do on Earth, each one can construct an
unbiased estimator of the $C_l$ for each observer position in the simulation
\begin{equation}
\widehat{C}_l(\bfx)=\sum_{m=-l}^l{|\alm(\bfx)|^2\over2l+1}
.\label{ClEstimator}
\end{equation}
and, assuming ergodicity, one should take a volume average of the estimators
from all the observers: 
\begin{equation}
\overline{\widehat{C}_l}
={1\over V}\int d^3\bfx\,\widehat{C}_l(\bfx)
=\sum_\bfk\sum_{m=-l}^l{|\Deltalm(\bfk)|^2\over2l+1}\ .
\label{ClVolumeEstimator}
\end{equation}
This is the best estimator one can obtain from a single simulation.  To obtain
the 2nd equality in this equation the orthonormality of the
$e^{i\bfk\ccdot\bfx}$ and the  sum rule for $D^l_{mm'}$'s has been used.  Since
only scalar, vector, and tensor modes contribute one may rewrite this as
\begin{equation}
\overline{\widehat{C}_l}=\overline{\widehat{C}_l^\rmS}
                        +\overline{\widehat{C}_l^\rmV}
                        +\overline{\widehat{C}_l^\rmT}
\label{ClDecomposition}
\end{equation}
where
\begin{eqnarray}
\overline{\widehat{C}_l^\rmS}&=&
                         \sum_\bfk{|\Delta_{\s(l, 0)}(\bfk)|^2\over2l+1} \cr
\overline{\widehat{C}_l^\rmV}&=&
                         \sum_\bfk{|\Delta_{\s(l,+1)}(\bfk)|^2
                                  +|\Delta_{\s(l,-1)}(\bfk)|^2\over2l+1} \cr
\overline{\widehat{C}_l^\rmT}&=&
                         \sum_\bfk{|\Delta_{\s(l,+2)}(\bfk)|^2
                                  +|\Delta_{\s(l,-2)}(\bfk)|^2\over2l+1} .
\label{ClScalarVectorTensor}
.\end{eqnarray}
Note that there are no cross-terms from the different helicities, just as there
is none for different $\bfk$-modes, which results from the fact that these
modes are orthogonal.  Note that if one only knows the pattern of anisotropy as
viewed from a given point one does not have enough information to decompose the
pattern one observes into it's scalar, vector, and tensor parts.  One can do so
only if one knows the pattern throughout space, which we do in the case of a
simulation.

\section{The Defects}

Usually at late enough epochs and large enough scales to be of interest for
CMBR anisotropy that the defects only interact with the rest of the matter
gravitationally. For example in the case of (non-superconducting) cosmic
strings the GUT scale strings are very thin and the ``scattering width''
extremely small by astronomical standards.  Since the gravitational field of an
object is completely determined by it's stress-energy tensor it is sufficient
to describe the evolution of defects by the time-history of their stress-energy
tensor, i.e. $\Theta_{\mu\nu}(\bfx,t)$.  This real symmetric matrix has ten
components, however in order for it to be compatible with General Relativity it
must obey energy and momentum conservation which reduces the number of
independent components to six.

	Defects are usually treated in the ``stiff source'' approximation which
basically means that the evolution of the defects are evolved in the metric of
the unperturbed universe.  The reasoning is that since any workable model of
defects can only produce small metric perturbations which implies that the
stress-energy of the defects must be small compared to that of the other matter
in the universe.  Taking into account the ``back-reaction'' of this small
perturbations on the stress-energy of the defect would be to include terms of
2nd order in smallness.  In mathematical language this means that one expands
everything to 1st order in $\Theta_{\mu\nu}$.  Strictly speaking it is true
that some defects are highly localized in space and can locally dominated the
stress-energy, but such defects also have enormous internal stresses, 
comparable to their energy density, and their motion is little effected by the
weak gravitational fields of the inhomogeneities over the course of a dynamical
time.  The stiff-source approximation does miss out on the cumulative effects
of back-reaction over many dynamical times, such as the decay of cosmic strings
into gravity waves, but these effects are usually only important on small
scales and do not have much direct effect on CMBR anisotropies.

	The stiff source approximation combined with linear perturbation theory
means that one can evolve the defects independently of the rest of the matter
and that the everything besides this evolution is linear.   The equations for
the evolution of the background matter are linear in $\Theta_{\mu\nu}$ so the
solution will consist of a term which is linear in $\Theta_{\mu\nu}$ and one
which depends on the initial conditions.  For example the CMBR anisotropy may
be written as
\begin{equation}
{\Delta T\over T}={\Delta T\over T}^\rmS+{\Delta T\over T}^\rmI
\label{SubsequentAndCompensation}
\end{equation}
where the 1st term is linear in $\Theta_{\mu\nu}$:
\begin{equation}
{\Delta T\over T}^\rmS(\hatbfn,\bfx,\eta)=\int_\etai^\eta d\eta'\,
\int d^3\bfx'\,D^{\mu\nu}(\hatbfn,\bfx-\bfx',\eta,\eta')\,
\Theta_{\mu\nu}(\bfx',\eta)
.\label{Subsequent}
\end{equation}
Here we have used conformal time $d\eta=dt/a$ where $a(t)$ is the cosmological
scale factor.  The $D^{\mu\nu}$ are given by solutions of the coupled
Einstein-matter equations of motion and will depend on the matter content of
the universe but not at all on the defects.  Causality assures us that
$D^{\mu\nu}=0$ when $|\bfx-\bfx'|>\eta-\eta'$.
The $(\Delta T/T)^\rmI$ term is often referred to as the {\it compensation}.
We should stress that this S/I decomposition as well as the exact form of
$D^{\mu\nu}$ will depend on how one chooses to solve the equations. In
particular by substituting any of the equations of energy-momentum conservation
into eq~(\ref{Subsequent}) and integrating by parts one would obtain a
different expression for $D^{\mu\nu}$ and one might also incorporate the
resulting boundary term on the initial hypersurface into the compensation,
obtaining a different expression for $(\Delta T/T)^\rmI$.  There is, however,
some physics in the compensation; namely the assumption that the universe was
completely homogeneous before the production of the defects; at least to the
extent that the inhomogeneities produced subsequently by the creation and
motion of the defects dominates over any initial inhomogeneity.  We do not
discuss compensation further as this is not a large part of the computation.

\subsection{Scalar-Vector-Tensor Decomposition}

	We have decomposed the brightness pattern into scalar, vector, and 
tensor parts, and one can do the same for both the stress-energy tensor and the
Green function $D^{\mu\nu}$.  For each $\bfk$-mode define a orthonormal basis
\begin{equation}
\hatbfk\quad \hatbfm_\bfk^| \quad \hatbfm_\bfk^- \quad {\rm where} \quad
  \hatbfk\equiv{\bfk\over|\bfk|} \ .
\label{WavenumberBasis}
\end{equation}
The orthonormality determines $\hatbfm_\bfk^|$ and $\hatbfm_\bfk^-$ up to a
rotation about $\hatbfk$, and we do not to specify it any further than that. We
may decompose the stress-energy tensor as
\begin{eqnarray}
\widetilde{\Theta}_{{\s0}i}&=&+\widetilde{\Theta}^\spar\hatbfk
                              +\widetilde{\Theta}^|    \hatbfm^|
                              +\widetilde{\Theta}^-    \hatbfm^-            \cr
\widetilde{\Theta}_{ij}&=&
 {1\over3}\widetilde{\Theta}^\circ\bfI
+\widetilde{\Theta}^{\spar\spar}(\hatbfk\hatbfk-{1\over3}\bfI)              \cr
&&+\widetilde{\Theta}^{\spar|}(\hatbfk  \hatbfm^|+\hatbfm^|\hatbfk  )
  +\widetilde{\Theta}^{\spar-}(\hatbfk  \hatbfm^-+\hatbfm^-\hatbfk  )       \cr
&&+\widetilde{\Theta}^+       (\hatbfm^|\hatbfm^|-\hatbfm^-\hatbfm^-)
  +\widetilde{\Theta}^{\times}(\hatbfm^|\hatbfm^-+\hatbfm^-\hatbfm^|)\ ,
\label{HelicityStressEnergy}
\end{eqnarray}
where we have used Latin indices, $i,j,k,\ldots$ to denote spatial coordinates.
Symmetry arguments tell us that the Fourier transform (wrt $\bfx-\bfx'$) of
$\widetilde{D}^{\mu\nu}$ may be decomposed as
\begin{eqnarray}
 \widetilde{D}^{{\s0}i}(\hatbfn,\bfk,\eta,\eta')&=&
 \widetilde{D}^\spar\,\hatbfk
+\widetilde{D}^\perp\,( \cos\phi_\rmk\hatbfm^|+\sin\phi_\rmk\hatbfm^-)      \cr
 \widetilde{D}^{ij}    (\hatbfn,\bfk,\eta,\eta')&=&
 \widetilde{D}^\circ       \,\bfI
+\widetilde{D}^{\spar\spar}\,(\hatbfk\hatbfk-{1\over3}\bfI)                 \cr
&&\hskip-50pt
+\widetilde{D}^{\spar\perp}\,
                      ( ( \hatbfk\hatbfm^|+\hatbfm^|\hatbfk)\cos\phi_\bfk
                       +( \hatbfm^-\hatbfk+\hatbfk\hatbfm^-)\sin\phi_\bfk)  \cr
&&\hskip-50pt
+\widetilde{D}^{\perp\perp}\,
     \left( (\hatbfm^|\hatbfm^|-\hatbfm^-\hatbfm^-)\cos2\phi_\bfk
           +(\hatbfm^|\hatbfm^-+\hatbfm^-\hatbfm^|)\sin2\phi_\bfk\right)    \cr
&&\hskip200pt
\label{GreenDecomposition}
\end{eqnarray}
where we use the notation 
\begin{equation}
\bfI=\delta_{ij} \qquad
\mu_\bfk\equiv\hatbfn\ccdot\hatbfk   \quad
\cos\phi_\bfk={\hatbfn\ccdot\hatbfm_\bfk^|\over\sqrt{1-\mu_\bfk^2}} \quad
\sin\phi_\bfk={\hatbfn\ccdot\hatbfm_\bfk^-\over\sqrt{1-\mu_\bfk^2}}\ .
\label{KmodeAngles}
\end{equation}
Here the arguments of $D^X$ are $(\mu_\bfk,|\bfk|,\eta,\eta')$, and there is
no dependence on $\phi_\bfk$.

\subsection{Anisotropy From the Defects}

	Combining eqs~(\ref{HelicityStressEnergy}\&\ref{GreenDecomposition}) we
find that eq~(\ref{Subsequent}) becomes
\begin{eqnarray}
{\Delta T\over T}^\rmS(\hatbfn,\bfx,\eta)=
\sum_\bfk e^{i\bfk\ccdot\bfx}\int_\etai^\etaf d\eta'
&&\hskip-22pt
\Biggl[
    \left(\widetilde{D}^{\s00}      \widetilde{\Theta}_{\s00}
         +\widetilde{D}^{\spar}     \widetilde{\Theta}^\spar
         +\widetilde{D}^{\circ}     \widetilde{\Theta}^\circ
+{2\over3}\widetilde{D}^{\spar\spar}\widetilde{\Theta}^{\spar\spar}\right)  \cr
&&\hskip-120pt
 +\left(  \widetilde{D}^{\perp}     \widetilde{\Theta}^|
        +2\widetilde{D}^{\spar\perp}\widetilde{\Theta}^{\spar|}\right)\,
                                                             \cos\phi_\bfk
 +\left(  \widetilde{D}^{\perp}     \widetilde{\Theta}^-
        +2\widetilde{D}^{\spar\perp}\widetilde{\Theta}^{\spar-}\right)\,
                                                             \sin\phi_\bfk  \cr
&&\hskip-22pt
 +       2\widetilde{D}^{\perp\perp}\widetilde{\Theta}^+     \,\cos2\phi_\bfk
 +       2\widetilde{D}^{\perp\perp}\widetilde{\Theta}^\times\,\sin2\phi_\bfk
       \Biggr]\ .\cr
&&\hskip100pt
\label{AnisotropyHelicity}
\end{eqnarray}
The above expressions for $(\Delta T/T)^\rmS$ is written so that the 1st, 2nd,
and 3rd lines contain only scalar, vector terms, and tensor terms,
respectively. Since this equation represents a purely gravitational effect it
contains no higher helicities.  If one compares eq~(\ref{AnisotropyHelicity})
to eq~(\ref{FourierHelicity}) and defines $\cos\theta_\bfk=\mu_\bfk$ one sees
that the angles $(\theta_\bfk,\phi_\bfk)$ are just the arguments of the
$\Ylm$'s in the helicity decomposition of the brightness pattern.  By equating
terms one see that
\begin{eqnarray}
&&\hskip-20pt
\Delta_{\s(l,0)}(\bfk)=i^l \sqrt{2\pi}\,
\int_\etai^\etaf d\eta'\,
               \left(D_l^{\s00}      \widetilde{\Theta}_{\s00}
                   +iD_l^{\spar}     \widetilde{\Theta}^\spar
                    +D_l^{\circ}     \widetilde{\Theta}^\circ
           +{2\over3}D_l^{\spar\spar}\widetilde{\Theta}^{\spar\spar}\right) \cr
&&\hskip-20pt
\Delta_{\s(l,\pm1)}(\bfk)=-i^l \sqrt{\pi\over2}
\int_\etai^\etaf d\eta'\,
    \left(  iD_l^{\perp}     \,(      \widetilde{\Theta}^|
                                \mp i \widetilde{\Theta}^-)
           +2D_l^{\spar\perp}\,(      \widetilde{\Theta}^{\spar|}
                                \mp i \widetilde{\Theta}^{\spar-})\right) \cr
&&\hskip-20pt
\Delta_{\s(l,\pm2)}(\bfk)=i^l\sqrt{2\pi}
\int_\etai^\etaf d\eta'\,D_l^{\perp\perp}\,(      \widetilde{\Theta}^+
                                            \mp i \widetilde{\Theta}^\times)
\label{HelicityIntegrals}
\end{eqnarray}
where we have expanded the $D$'s in terms of spherical harmonics:
\begin{equation}
\widetilde{D}^X
=[i](-1)^m\sqrt{2\pi}\,\sum_{l=0}^\infty i^l\,\Ylm(\theta_\bfk,0)\,
                                         \widetilde{D}^X_l
\qquad m=\left\{\matrix{ 0 & \scriptstyle{X=00,\spar,0,\spar\spar} \cr
                         1 & \scriptstyle{X=\perp,\spar\perp}      \cr
                         2 & \scriptstyle{X=\perp\perp}               }\right.
\end{equation}
The extra factor, $[i]$, is only for $\widetilde{D}^\parallel$ and
$\widetilde{D}^\perp$ since with this definition all the $\widetilde{D}^X_l$'s
are real.

	We see from the above that there are at most $7$ Green functions which
one must compute, 4 scalar ($D^{\s00}$, $D^\spar$, $D^{\s0}$,
$D^{\spar\spar}$), 2 vector  ($D^\perp$, $D^{\spar\perp}$), and one tensor
($D^{\perp\perp})$. However one can use the equations of energy-momentum
conservation, which may be written
\begin{eqnarray}
\dot{\widetilde{\Theta}}_{\s00}+
 {\dot{a}\over a}(\widetilde{\Theta}_{\s00}+\widetilde{\Theta}^\circ)
                                      =ik \widetilde{\Theta}^\spar   &\qquad&
\dot{\widetilde{\Theta}^\spar}+2{\dot{a}\over a}\widetilde{\Theta}^\spar
  =ik(\widetilde{\Theta}^\circ+{2\over3}\widetilde{\Theta}^{\spar\spar}) \cr
\dot{\widetilde{\Theta}^-}+2{\dot{a}\over a}\widetilde{\Theta}^-
                                      =ik\widetilde{\Theta}^{\spar-} &\qquad&
\dot{\widetilde{\Theta}^|}+2{\dot{a}\over a}\widetilde{\Theta}^|
                                      =ik\widetilde{\Theta}^{\spar|}
\label{EnergyMomentumConservation}
\end{eqnarray}
to eliminate 2 of the scalar terms, and 1 (for each of $|$ and $-$) of the
vector terms.  Thus one really only needs to calculate 4 Green functions, 2
scalar, 1 vector, and 1 tensor.  The vector and tensor Green functions are each
used twice, once for each ``polarization'' of vector and tensor modes.

\subsection{Computational Cost}

	Now consider the cost to compute the full anisotropy pattern from a
simulation of the defects.  We do not include here the cost of simulating the
defects, i.e. producing the $\Theta_{\mu\nu}(\bfx,\eta)$, which typically
scales as $N^4$.  For a cubical simulation box the Fourier spectrum take on the
discrete values
\begin{equation}
\bfk={2\pi\over L}\bfN
\end{equation}
where $L$ is the size of the cube, and the components of the vector $\bfN$ is
any triplet of integers.  With $N^3$ spatial resolution elements there are
$N^3$ modes and the components of $\bfN$ take on the values
$\{-N+1,\,-N+2,\ldots,\,N-1,\,N\}$.  The Green functions depend only on
$|\bfk|$ which is $2\pi/L$ times the sum of the squares of 3 integers in this
range. Clearly there are no more than $3(N/2)^2$ such values.  For a million
different $\bfk$'s there are only a few thousand different $|\bfk|$'s.  Usually
one would want to compute these Green functions for one fixed observation time,
$\eta$, but would need to have temporal resolution in $\eta'$ the same as in
the simulation.  The Courant condition suggest that this resolution must
correspond to at least $N$ timesteps during the simulation.  Since the Green
functions are usually small if $l\gg k(\eta-\eta')$ we may {\it a priori}
ignore terms with larger $l$. Since the vast majority of modes have $k\sim
N/L$, and most of the required terms have $(\eta-\eta')\simgt L$; we see that
typically we want $D_l$'s for $l$ up to $\lmax\sim N$.  Thus a complete table
of Green functions would have size $\sim N^4$. One could certainly reduce this
greatly since, in principle, one needs a wavenumber resolution no better than
$\Delta k\sim 2\pi/L$, reducing the number of $k$-modes to something of order
$N$ rather than $N^2$.  To compute each of these Green functions numerically 
requires in integration from $\eta'$ to $\eta$, taking roughly $N$ operations;
so the cost of computing these Green functions scales like $N^4$ if one
does not oversample in $k$ ($N^5$ if one does).

	Performing the integrals of eq~(\ref{HelicityIntegrals}) is liable to
be a larger task, as this clearly scales as $N^5$; there being $N^3$ different
$\bfk$'s and $5\lmax$ integrals for every $\bfk$ and the integrals themselves
requiring $\sim N$ operations.  We do stress that the prefactors multiplying
these scaling laws can be reduced significantly by not computing every $l$ or
coarsening the resolution of the integrations.  Once one has done the integrals
one has the $\Deltalm$'s from which one could compute the best estimates of the
angular power spectrum using eq~(\ref{ClVolumeEstimator}), which is itself is
an $N^4$ computation.  However the power spectrum is not the entire story and
one would really like to have the full anisotropy pattern for of the observers
in the simulation volume.  To do this one needs to perform the Wigner rotation
of eq~(\ref{WignerRotation}), which converts the $\sim N^4$ $\Deltalm$'s to the
$\sim N^5$ $\tildealm$'s and is an $N^5$ computation. Finally to transform the
$\tildealm$'s to $\alm$'s one may use an FFT which is again an $N^5$
computation.  Note that if one does want to know the full brightness
distribution then the final data set is $\sim N^5$ in size and the
computational cost scales in the same way.  Clearly it could not scale in any
better way.

\section{Computational Reduction}

	We have seen that the cost of computing the full brightness
distribution from a defect simulation scales as $N^5$, which is more costly
than the cost of simulating the defects ($N^4$), and in any case makes the
computation large.  How can one can get away with doing less?  For computations
of large-angle anisotropy the effects of scattering is not important and the
anisotropy in defect models is largely determined by the Sachs-Wolfe integral
along the line-of-sight, perhaps with the inclusion of a boundary term at the
last-scattering surface.  In this case the brightness along a given
line-of-sight is independent of the brightness along other lines-of-sight and
one may compute these separately if one knows the metric perturbation
throughout space-time (an $N^4$ computation).  Thus one can concentrate all
ones effort on the photons converging on a small number of observers
\cite{BBS88,BennettRhie93,DHZ93,PST94,PiInSky94,PS95,DZ96,Turok96,ACSSV96},
performing an $N^3$ computation for each 
observer.\footnote{Bouchet \etal\cite{BBS88} and Allen \etal\cite{ACSSV96} use
a somewhat different technique which avoids computing the metric perturbation
directly,and is particularly advantageous for cosmic strings which do not fill
space.}

	When scattering is important, as for small-angle anisotropy, the
brightness along a given line-of-sight will, strictly speaking, depend on the
brightness in all other directions at all points along that lines-of-sight.
Since the number of these brightnesses is $\sim N^5$, one might worry that one
requires an $N^5$ computation to find the anisotropy pattern for just one
observer! However in practice one finds that this dependence is almost
exclusively on the first few moments of the brightness pattern (e.g. monopole
and dipole) and depends only very weakly on the higher moments, and that one
can compute these lower moments without much regard to the higher moments.
These sorts of approximations have been in use for 
decades\cite{Dautcourt,DavisBoynton} they have been
perfected only recently\cite{HS96,SZ96}.  In practice computing the anisotropy
with scattering is no larger a computation than without, one need only do a few
integrals along the line of sight!  Note however that if one were to attempt
computing the anisotropy pattern in this way for all $N^3$ observers in the box
one would end up with a computation which scales as $\sim N^6$.  Given that the
results for different observers from the same simulation are not completely
statistically independent it is not necessarily worthwhile to compute the
anisotropy pattern for all observers allowed by ones spatial resolution.
Allen~\etal\cite{ACSSV96} have used 64 observers in a single large-angle
simulation and found no significant correlations between observers.  Given that
the cost per observer in a given simulation goes as $N^2$ while the cost of a
simulation goes as $N^4$ it is certainly worthwhile to use many observers in
a given simulation.

\section{Another Algorithm}

	When there is scattering one may restrict oneself to a few
observers but still use the Fourier representation described above.  The
classic use\cite{BBS88,PiInSky94,Turok96} of a small-angle simulation is to
only compute the brightness for lines-of-sight traveling along one or more of
the 6 principle axes of the box ($\pm\hatbfx$, $\pm\hatbfy$, $\pm\hatbfz$), as
illustrated in fig~\ref{fig:SmallAngle}.  To do this one need just use the
appropriate values of $\mu_\bfk$ and $\phi_\bfk$ in
eq~(\ref{AnisotropyHelicity}) for the direction one is interested in. What one
obtains is the brightness pattern, $\Delta T/T(\hatbfn,\bfx)$, but with fixed
$\hatbfn$.  As long as $\hatbfn$ is along one of the principal axes then
each slice of this function of $\bfx$ perpendicular to $\hatbfn$ gives the
brightness pattern which would be seen by a ``distant'' observer on a square
patch of sky.  There would be $N$ different slices which  represent the
sequence of patterns a given observer would see over a period of time.
Performing the integrals is an $\sim N^4$ algorithm (FFTing to real space only
$\sim N^3$) so the cost per pattern is $\sim N^3$, the same as required for a
similar result using the line-of-sight method mentioned above.  Of course there
will be some correlations between the patterns on nearby slices.

\section{Summary}

	This report lists some of the mathematical details of CMBR anisotropy
which could be used in computations of CMBR anisotropy from defects.  Hopefully
these details are interesting even to those not planning to do the
calculations.  Many other details were not mentioned for lack of space,
specifically those relating to computing the Green functions.  The equations 
used are just the usual cosmological Boltzmann-Einstein equations with the
appropriate initial conditions.  The formalism specifically outlined has not
been used in the past (which is why it was concentrated on) but some papers
applying some of these technique to cosmic string simulations will appear
soon\cite{ACDKSS}.

\section*{Acknowledgments} This work was supported by the DOE and NASA grant
NAG5-2788 at Fermilab.

\end{document}